\documentclass[conference]{IEEEtran}
\usepackage{cite}

\ifCLASSINFOpdf
  \usepackage[pdftex]{graphicx}
   \graphicspath{{C:/Users/Administrator/Desktop/latex/eps}}
  \DeclareGraphicsExtensions{.pdf,.jpeg,.png,.eps}
\else
\fi
\usepackage{CJK}
\usepackage[numbers,sort&compress]{natbib}
\usepackage{cite}
\usepackage{amsmath,amssymb,enumerate}
\usepackage{cite,graphics,graphicx}
\usepackage{graphicx}
\usepackage{subfigure}
\usepackage{listings}
\usepackage{makeidx}

\usepackage{algorithm}
\usepackage{algorithmic}
\usepackage{float}
\usepackage{multirow}
\usepackage{amsmath}
\usepackage{hyperref}
\usepackage{amssymb}
\usepackage{xcolor}
\usepackage{epsfig}
\usepackage{graphics}
\usepackage{listings}
\usepackage{cite}
 \usepackage{listings}
\usepackage{algorithm}
\usepackage{algorithmic}
\hypersetup{linkcolor=black, %
            citecolor=black, %
             pdftitle={Title}, %
            pdfauthor={Name}, %
           pdfsubject={Subject}, %
          pdfkeywords={Key words}}
\begin{document}

%
\title{A Way For Accelerating The DNA Sequence Reconstruction Problem By CUDA}

\author{\IEEEauthorblockN{Yukun Zhong{$^{1}$},
ZhiWei He{$^{2}$},
XianHong Wang{$^{3}$},
XiongBin Cao{$^{4}$}}
\IEEEauthorblockA{
Computer Science and Engineering Department\\ Sichuan University Jinjiang College, Penshan 620860, China\\
Email: dreamertimer7@gmail.com}
}

%


\maketitle

\begin{abstract}
Traditionally, we usually utilize the method of shotgun to cut a DNA
sequence into pieces and we have to reconstruct the original DNA sequence from the
pieces, those are widely used method for DNA assembly. Emerging DNA sequence technologies open up more
opportunities for molecular biology. This paper introduce a new method to improve the efficiency of reconstructing DNA sequence using suffix array based on CUDA programming model. The experimental result show the construction of suffix array using GPU is an more efficient approach on Intel(R) Core(TM) i3-3110K quad-core and
NVIDIA GeForce 610M GPU, and study show the performance of our method is more than 20 times than
that of CPU serial implementation. We believe our method give a cost-efficient solution to the
bioinformatics community.\\\\
keywords-CUDA, DNA Sequence assembly, GPU, shotgun method , Supersring, Suffix Array, Radix Sort
\end{abstract}


%
\IEEEpeerreviewmaketitle

\section{Introduction}
 Recent advances in molecular biology techniques such as automated DNA sequencing
require scientists to promptly gather huge amounts of gene sequence and expression data.
Due to such a heavy amount of data, the projects of genes become a time-consuming and challenging
problem, and the development of bioinformatics required to reach
an optimal performance. Several efforts have been made to
increase the speed or to reduce the computational
cost. The shotgun sequencing strategy has been used widely in genome sequencing projects. In this thesis, we shall state the DNA assembly problem with data obtained by the shotgun approach is special case, and also show that the shortest common superstring is not necessarily the correct solution. The data of the DNA assembly problem that we present is shown to be special and satisfy some special condition. The modified method first pointed by J. P. Lu, “DNA Sequence Assembly,”\cite{Lu:Thesis:2004}. In computational methods, biological
sequences are viewed as strings and this paper introduce
the suffix array which is a widely used method for bioinformatics
. The method that we present to accelerate the DNA reconstruction is searching by suffix arrays based on CUDA. \\\\

Now CPU microprocessor based on a
single central processing promote the performance of the
computer application and reduce cost, making the floating point
arithmetic achieved 11 times per second single chip, the era
of the many-core processor has begun .\cite{kirk2010programming} The emergence
of many-core architectures, such as compute unified device
architecture(CUDA)-enabled GPUs.\cite{anderson2008general}. Life science have emerged as a primary application area
for the use of GPU computing. High-throughput techniques
for DNA sequencing and gene expression analysis have led to a data explosion.
Prominent examples are the growth of
DNA sequence information in NCBI’s GenBank database and
the growth of protein sequences in the UniProtKB/TrEMBL
database. Due to the exponentially growing of DNA bases, the development of bioinformatics need to reach
an optimal performance. Several efforts have been made to
increase the convergence speed or to reduce the computational
cost. Suffix array which is naturally considered
be high parallel is constructed by CUDA, and the implementation shows the solving reconstruction of DNA problem on CUDA is a good and
efficient method.  \\\\

\section{general introduction of CUDA}
 CUDA (Compute Unified Device Architecture) is
an extension of C/C++ to write scalable multi-threaded
programs for CUDA-enabled GPUs.\cite{ryoo2008optimization} CUDA is a parallel computing platform and programming
model invented by NVIDIA. It enables dramatic increases in
computing performance by harnessing the power of the graphics processing unit (GPU)\cite{nvidia2007compute}\cite{kirk2010programming}. The GPU is treated as a coprocessor
that executes data-parallel kernel code. The user supplies a single
source program encompassing both host (CPU) and kernel (GPU)
code. Since yielded in 2007, hundreds of millions
of computers equipped with CUDA processing capability
of GPU is widely used by people, CUDA to users through
a widely used inexpensive hardware performance provides
a good opportunity to reduce the floating point arithmetic
or other numerical computation time.\cite{ryoo2008optimization} GPU parallel technology is now widely used in various fields, the GPU high-speed parallel
processing capabilities could be used in digital image
processing algorithms,discrete simulation, general computing,
greatly improving the computational speed, because the
GPU performance characteristics can be applied to numerical
calculation and matrix processing.\cite{wen2011gpu}\\\\
\begin{figure}[!t]
\centering
\graphicspath{{C:/Users/Administrator/Desktop/latex 格式/IEEE_CS_Latex8.5x11x2}}
\includegraphics[width=2.5in]{hhf.pdf}
 \DeclareGraphicsExtensions.
\caption{CUDA Architecture}
\label{fig 1}
\end{figure}
  The CUDA programming model is very well suited to expose the parallel
capabilities of GPUs. The latest generation of NVIDIA GPUs, based on the Tesla
architecture, supports the
CUDA programming model and tremendously accelerates CUDA applications. CUDA threads may access data from multiple memory spaces during their
execution as illustrated in fig\ref{fig 1}.Each thread has a private local memory. Each
thread block has a shared memory visible to all threads of the block and with the
same lifetime as the block. Finally, all threads have access to the same global
memory.
\section{DNA Sequence Reconstruction and Related algorithm}
 In molecular biology filed, we often utilize the method of shotgun to cut a DNA sequence into pieces and we have to reconstruct the original DNA sequence from the pieces. This process is often practiced because a DNA sequence is a long one and we have to cut it before we can study it. \cite{huang1999cap3}Reconstructing the DNA sequence from a set of substrings efficiently becomes important and tough problem.In this paper, we shall first point out that the DNA assembly problem with data obtained by the shotgun approach is special in the sense that there may be several feasible solutions and we have no way to know which one is the original one. A DNA sequence is represented by a string of characters drawn
from a four-letter alphabet (A, C, G, and T) corresponding to the
four monomeric bases of which the DNA polymer is composed.
A piece, or fragment, corresponds in our context to a sub-string \cite{burks1994dna}.
 In the thesis, we mainly give a  DNA assembly problem with data obtained by the shotgun approach which is special in the sense.The tradition problem can be defined as :
 given a  string D $=(d1 ,d2 , d3 , d4..)$ which are obtained from an original string by using shotgun method, the main work is to reconstruct the original string from the D.
 \begin{figure}[H]
\centering
\graphicspath{{C:/Users/Administrator/Desktop/latex 格式/IEEE_CS_Latex8.5x11x2}}
\includegraphics[width=2.5in]{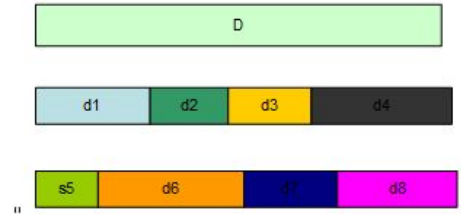}
 \DeclareGraphicsExtensions.
\caption{DNA Cutting segment}
\label{fig 2}
\end{figure}
 In the picture \ref{fig 2} depict the sequence cut into m segment by the first shotgun. The picture demonstrate the second cutting.  Now assume we just obtain two sets of segment and input the segment into only one group which contain the two segments, the problem will be $(m+n)$ segment. As given for example :\\
  assuming we have a string :\\\\ ${abthatb, hatbpaab, tbabhhatbpaa, paabtabh, bhaabtpb}$\\\\ In order to reconstruct the original sequence, there are many methods proposed, For example, Gallant et al. Have shown that the determination of the shortest common superstring (SCS) for a fourletter alphabet, in which the fragment assembly problem can be recast, is NP-complete (i.e., computationally infeasible for large data sets) [Gallant et al., 1980].
Even the over-simplified, reduced approach of modeling fragment assembly as determining the order of
beads (fragments) on a string is NP-complete \cite{parsons1993genetic}. Now people are using a method called " superstring " to deal with such problem, so we can get the weighted directed graph ：\\\\
\begin{figure}[H]
\centering
\graphicspath{{C:/Users/Administrator/Desktop/latex 格式/IEEE_CS_Latex8.5x11x2}}
\includegraphics[width=2.5in]{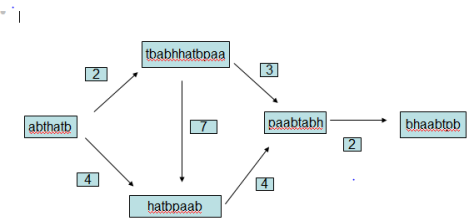}
 \DeclareGraphicsExtensions.
\caption{weighted directed graph}
\label{fig 3}
\end{figure}
 in fig\ref{fig 3}, in the picture we get maximal Hamiltonian path :\\\\
 ${abthatb\Rightarrow  tbabhhatbpaa\Rightarrow hatbpaab \Rightarrow paabtabh\Rightarrow bhaabtpb }$\\\\

 As followed, we can obtain the sequence :\\\\
 ${abthatbabhhatbpaabtabhaabtpb}$\\\\
  this is also the shortest common superstring \\${abthatbabhhatbpaabtabhaabtpb}$ \\of the input string . The superstring problem was proved to be an NP-hard. Obviously, we must follow the rule :\\\\
  DEFINITION :  The DNA sequence reconstruction problem , given an error rate $ 0 \le \mu < 1 $ and F of fragment sequences, find shortest sequence S such that for every fragment A $\in$ F , there is a substring B such that\cite{kececioglu1995combinatorial} \\\\
  min （(A,B),($\bar{A},B)$）$\le$$\mu$$|$A$|$\\\\
  But in some cases , the situation is different .\\\\
  \begin{figure}[H]
\centering
\graphicspath{{C:/Users/Administrator/Desktop/latex 格式/IEEE_CS_Latex8.5x11x2}}
\includegraphics[width=2.5in]{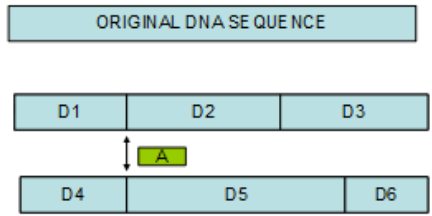}
 \DeclareGraphicsExtensions.
\caption{special situation}
\label{fig 4}
\end{figure}
  In picture \ref{fig 4}, we can easily find if two segment cut in a same position, it cannot manage to reconstruct the original DNA sequence. In this case, we must suppose all breaking position are different. So this paper do not introduce the DNA sequence assembly as the shortest superstring problem.First, we called a fragment $D_{k}$ 'fir' when the $D_{k}$ is prefix of another fragment or there is a fragment$D_{i}$ being the prefix of $D_{k}$. The basic idea would follow the process as depicted :
\begin{figure}[H]
\centering
\graphicspath{{C:/Users/Administrator/Desktop/latex 格式/IEEE_CS_Latex8.5x11x2}}
\includegraphics[width=2.5in]{hgg.pdf}
 \DeclareGraphicsExtensions.
\caption{DNA sequence reconstruction}
\label{fig 5}
\end{figure}
we can easily find that both of $d_{1}$ and $d_{3}$ are "fir", in other wordS, the $d_{3}$ is prefix of $d_{1}$. In the case, it notes that if one fragment and another are both fir fragment, especially one of them is not very short,sometimes we can assert it has high probability that both of them are the beginning fragment. After finding the $d_{3}$ is the prefix of the $d_{1}$ , we cut the $d_{3}$ from the prefix $d_{1}$, so we can get the remain we called "$d^{'}_{3}$",depicted :
\begin{figure}[H]
\centering
\graphicspath{{C:/Users/Administrator/Desktop/latex 格式/IEEE_CS_Latex8.5x11x2}}
\includegraphics[width=2.5in]{bb.pdf}
 \DeclareGraphicsExtensions.
\caption{DNA sequence reconstruction}
\label{fig 6}
\end{figure}
 and the remain fragment can induce we to next process as we did.
\begin{figure}[H]
\centering
\graphicspath{{C:/Users/Administrator/Desktop/latex 格式/IEEE_CS_Latex8.5x11x2}}
\includegraphics[width=2.5in]{qq.pdf}
 \DeclareGraphicsExtensions.
\caption{DNA sequence reconstruction}
\label{fig 7}
\end{figure}
 This process is repeated again and again until we have the get situation:
It is notable that sometimes the fragments we choose may not be what we want, so in this method it let RS be the result of the concatenation of the selected fragments. If a fragment is neither selected nor marked by this method , it also illustrate the RS is not a superstring of the input fragments.In a same way,the RS won't be the correct reconstructed
sequence. In order to deal with the situation, the method must use the back-track to reach the condition we expect. Observation in the picture:
\begin{figure}[H]
\centering
\graphicspath{{C:/Users/Administrator/Desktop/latex 格式/IEEE_CS_Latex8.5x11x2}}
\includegraphics[width=2.5in]{uu.pdf}
 \DeclareGraphicsExtensions.
\caption{DNA sequence reconstruction}
\label{fig 7}
\end{figure}
the picture show a situation we always want to avoid, in this case there are obvious three fir fragments which are $d_{1}$, $d_{4}$, $d_{6}$. We can simple find $d_{1}$ is the prefix of $d_{6}$ and $d_{4}$, but how could we do next.
suppose we choose the $d_{1}$ and $d_{6}$, we cut the  $d_{1}$ segment from the $d_{6}$ and we can obtain new fragment $d_{6}$, it is notable that there is no fragment in the sequence after the cutting. So the back-tracking is need to solve the problem, the next step we should choose the others $d_{1}$ and $d_{4}$, now remove the $d_{1}$ segment from the fragment $d_{4}$, the remain segment of $d_{4}$ make process continued. We give another situation which is not satisfied to explain how terrible these cases are. Look this case shown :
 \begin{figure}[H]
\centering
\graphicspath{{C:/Users/Administrator/Desktop/latex 格式/IEEE_CS_Latex8.5x11x2}}
\includegraphics[width=2.5in]{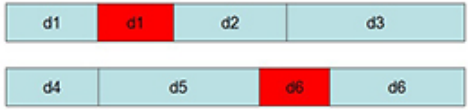}
 \DeclareGraphicsExtensions.
\caption{DNA sequence reconstruction}
\label{fig 7}
\end{figure}
It is so clear find the $d_{1}$ and $d_{4}$ are fir. First we cut remove the $d_{4}$ segment from the fragment $d_{1}$ and then we will find the $d_{1}$ in red is the prefix of $d_{5}$ and $d_{6}$, this time we must choose two fragment to eliminate the redundant segment. Suppose we select the $d_{1}$ and $d_{6}$, process will terminate at this step, because there is no fragments whose are fir. If we change a little in the picture, we turn the segment $d_{6}$ in blue into segment $d_{7}$. Now some mistakes might happen, after the first step which remove the segment $d_{4}$ from the segment $d_{1}$, it is not difficult for us to find three fragment are fir ($d_{1}$, $d_{5}$ and $d_{6}$). Let $d_{1}$ and $d_{6}$ into the execution flow first, it obtains a wrong solution. Because $d_{1}$ is equal to $d_{6}$, we regard the segment $d_{6}$ as last fragment on the basis of the method. So the fragment $d_{2}$ $d_{3}$ $d_{5}$ $d_{7}$ are unselected fragment. Given three cases to demonstrate the back-tracking is need when the wrong situation occurs.
\section{parallel method for reconstruction by searching in suffix array}
 This method we proposed has been proved to be a good and heuristic algorithm, but the experiment shows the algorithm has disadvantage of execution efficiency . But we can find the problem that it always need back-tracking when the process reach a wrong situation and it is obviously time-consuming job with increasing number of DNA sequence. As far as we know, Backtracking method is a kind of heuristic solving methods. If the test is successful, namely it gets solution; If the test fails, it shall gradually back, further testing in other route. But with the growing demanding of The Human Genome project, we need high efficiency of method to meet the demand of development. \\\\

    In this paper we introduce a method to construct suffix array based on GPU for improving the efficiency of the search of ID,when back-tracking happen. The suffix array is a space-efficient data structure that allows
efficient searching of a text for any given pattern. The suffix
array is basically a sorted array Pos of all the suffixes of a
text.\cite{satish2009designing} So it is easy and efficient for suffix array for binary search to solve the ID matching testing whether one of fragment in D is the prefix of X or X is the prefix of one in D. Traditional method build suffix array can be in O(nlog(n)). Our method to construct the suffix array using radix sort algorithm  is a simple yet very
efficient sorting method that outperforms many well known
comparison-based algorithms for a certain type of keys such
as integer. The sorting algorithm used within each pass of radix sort is typically a counting sort or bucket sort. To
compute the output index at which the element should be
written―which we will refer to as the rank of the element.
we must simply count the number of elements in lower
numbered buckets plus the number of elements already in
the current bucket.\cite{satish2009designing}
\\\\

\begin{figure}[H]
\centering
\graphicspath{{C:/Users/Administrator/Desktop/latex 格式/IEEE_CS_Latex8.5x11x2}}
\includegraphics[width=2.5in]{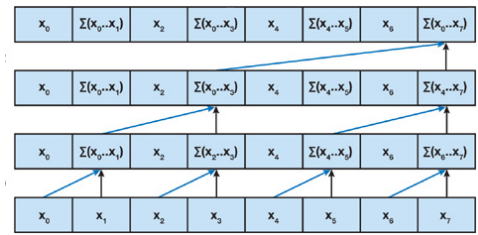}
 \DeclareGraphicsExtensions.
\caption{Scan to radx sort}
\label{fig dscan}
\end{figure}

  Now CUDA is considered to be a good way to drastically reduce the runtime. CUDA has their own advantage over other platform. it has its own
characteristics.For example if we use CUDA global memory, its memory large storage but speed of read and write is slow,
so we are going to use Shared memory. The main abstractions on which CUDA is
based are the notion of a kernel function, which is a single
routine that is invoked concurrently across many thread
instances;\cite{sintorn2008fast} A software controlled scratchpad, which CUDA
calls the “shared memory”, in each SIMD core and barrier
synchronization. CUDA presents a virtual machine consisting of an arbitrary number of streaming multiprocessors
(SMs), which appear as 32-wide SIMD cores with a total
of up to 512 thread contexts (organized into warps of 32
threads each).\cite{boyer2008automated}. This approach to implementing radix sort is conceptually quite simple. Given scan and permute primitives, it is
straightforward to implement illustrate in \ref{fig dscan}\\\\

\begin{figure}[H]
\centering
\graphicspath{{C:/Users/Administrator/Desktop/latex 格式/IEEE_CS_Latex8.5x11x2}}
\includegraphics[width=2.5in]{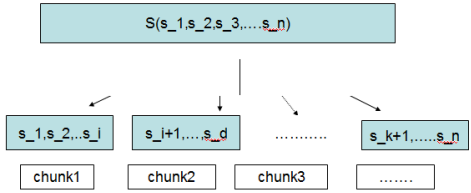}
 \DeclareGraphicsExtensions.
\caption{chunks of divided S}
\label{fig 17}
\end{figure}

 Parallelizing radix sort approach is also
based on dividing the sequence into blocks that can be processed by independent processors. This algorithm uses radix sort to sort individual chunks
of the input array. Chunks are sorted in parallel by multiple
thread blocks. Chunks are as large as can fit into the shared
memory of a single multiprocessor on the GPU.  Chunks are as large as can fit into the shared memory of a single multiprocessor on the GPU. Given a string $S=\lbrace s_{1}s_{2}s_{3}..s_{n}\rbrace$, divide S into every single chunk like $chunk_{1}=\lbrace s_{1}s_{2}..s_{i}\rbrace$, $chunk_{2}=\lbrace s_{i+1}s_{i+2}..s_{d}\rbrace$, ..$chunk_{..}=\lbrace s_{..},..,s_{n}\rbrace$.illustrate in \ref{fig 17}
 . Thus for k-bit keys, radix sort requires k steps. Our implementation requires one scan per step. The fundamental primitive we use to implement each step of radix sort is the split primitive. The input to split is a list of sort keys and their bit value b of interest on this step, either a true or false. \cite{sintorn2008fast}The output is a new list of sort keys, with all false sort keys packed before all true sort keys. With split, we can easily implement radix sort. We then initialize our current bit to the least-significant bit of the key, split based on the key, check if the output is sorted, and if not shift the current bit left by one and iterate again. When we are done, we copy the sorted data back to global memory. With large inputs, each chunk is mapped to a thread block and runs in parallel with the other chunks . Implementation of each pass of the radix sort for the i-th least
significant digit using four separate CUDA kernels.\cite{karkkainen2006linear}\\

1 . Sort each block in on-chip memory according to the
i-th digit using the split primitive .\\

2. Compute offsets for each of the r buckets, storing them
to global memory in column-major order .\\

3. Perform a prefix sum over this table.\\

4. Compute the output location for each element using
the prefix sum results and scatter the elements to their
computed locations\\

After each block-size chunk is sorted, we use a recursive
merge sort to combine two sorted chunks into one sorted
chunk. The the pseudocode is depicted in Alg.1 is In the
kernel function cudaDC3RS. In the kernel function id is a
private register with respect to a thread respectively, which
denote as thread index.Follow the Alg.1 all
single block of subsequence are organized into a string array
S ,so that the exclusive thread can be mapped into S[id]. Then
each subsequence is executed respectively so as to the indexed
structure enable all threads to string matching simultaneously
by means of radix sort algorithm supported by GPU.

 \begin{figure}[H]
\centering
\graphicspath{{C:/Users/Administrator/Desktop/latex 格式/IEEE_CS_Latex8.5x11x2}}
\includegraphics[width=2.5in]{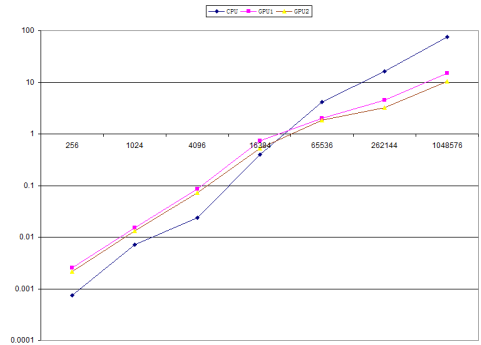}
 \DeclareGraphicsExtensions.
\caption{performance of CPU and kernel execution of GPU}
\label{fig 11}
\end{figure}

 \begin{figure}[H]
\centering
\graphicspath{{C:/Users/Administrator/Desktop/latex 格式/IEEE_CS_Latex8.5x11x2}}
\includegraphics[width=2.5in]{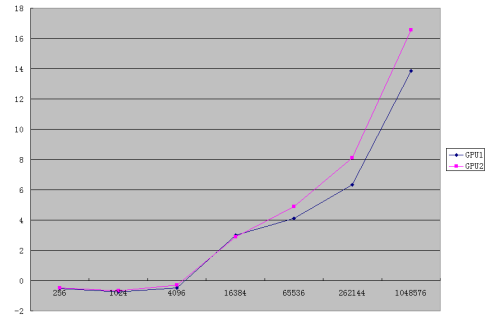}
 \DeclareGraphicsExtensions.
\caption{average speedup ratio between GPU and CPU}
\label{fig 12}
\end{figure}.

\section{experiment and result}
  The DNA sequence reconstruction performance executed on Intel(R) Core(TM) i3-3110K quad-core running at 2.40GHZ  with 2.0GB RAM, and NVIDIA GeForce 610M GPU(kepler architecture) which has a total of 48 streaming
multiprocessors operating at a clock rate of 900 MHZ. We obtained DNA sequence from NCBI database(Bacillus megaterium MSP20.1 sequence) to testify the method as we described. And several sets of Ds sequence ranging from 1024 to 1048576 were implemented in the experiment. a kernel and organize its execution in a grid of thread
blocks, Each block is assigned to a Streaming Multiprocessor
(SM). For each SM task allocation unit can allocate up to eight blocks. Each warp is made up of 32 threads. SM is one of the four instruction is minimum delay instruction cycle. Eight SP adopted launch an instruction, the implementation of four line structure. So Warp composed of 32 threads are the smallest units of CUDA program execution, and the same Warp is strictly serial, therefore is no synchronization in the Warp. And if the number of thread in each block  is not a multiple of 32, that it will put the rest of the thread into a warp independently; For example, thread number is 66, there will be three warp: 0-31, 32-63, 64-63. Due to there are two needless threads in a warp, so in calculation, is equivalent to waste the computing power of the 30 thread; This is in setting the thread in the block number must be to noticed In the \ref{fig 11} depicts the difference performance of different allocation of threads. The line of GPU1 stands for the number of threads in each block is not multiple of 32, GPU2 stands for the number is multiple of 32.\\\\
\begin{algorithm}[H]         
\caption{ KERNEL FUNCTION FOR CONSTRUCTION OF THE SUFFIX ARRAY }             
\label{alg:SA}                  
\begin{algorithmic}[1]                
\WHILE {each $i < 32$}
  \STATE  $row=blockIdx.x*twidth+threadIdx.x$;
  \STATE  $col=blockIdx.y*twidth+threadIdx.y$;
  \STATE   $id=row*width+col$;
  \STATE  $ set the index for each threads$
  \STATE  $b[id]=( S[id] \gg i)\&1$;
  \STATE   ${//}$ record i{\_}th bit of each element
  \STATE  ${\_\_}syncthreads()$;
   \STATE  $  e[id]=b[id]\otimes1$;
   \STATE  $ {\_\_}syncthreads()$;
  \STATE $temp[id] = e[id]$;
\STATE ${\_\_}syncthreads()$;
\FORALL{$oFF=1 $ , \textbf{and},$oFF<n$}
\STATE ${//}$scan the 1s table to accaculate prefix sum of each element
\STATE $oFF\ll 1$;
\IF{$id \le oFF$}
\STATE $temp[id]=temp[id-oFF]+temp[id]$;
\ELSE
\STATE$ temp[id] = temp[id]$;
\ENDIF
\STATE ${\_\_}synctreads()$;
\ENDFOR
\STATE $f[id]=temp[id]$;
\STATE  ${\_\_}syncthreads()$;
\STATE $ tof=e[n-1]+f[n-1]$
 \STATE $t[id]=id-f[id]+tof$;
 \STATE${\_\_}syncthreads()$;
 \STATE$d[id]=b[id]?t[id]:f[id]$;
 \STATE ${//}$array d store the index of the sorted S
 \STATE${\_\_}syncthreads()$;
\STATE $OutS[d[id]]=S[id]$;
\STATE $S=OutS$;
  \ENDWHILE
\end{algorithmic}
\end{algorithm}

 In previous experiment, the most disadvantage in the algorithm is the high cost of run-time in the implementation with massive amount of data. the CUDA designed for deal with massive data parallel computing problem now is used to implement the DNA sequence reconstruction, in the picture \ref{fig 11} show that the difference performance between  CPU and GPU performance  and in picture \ref{fig 12} show the rate speedup beteen the execution on CUDA architecture and series computing on Microsoft Visual Studio 2010 by CPU. Compared with CPU implementation which is executed on Microsoft Visual Studio 2010, we can easily find the tremendous distinction between CPU and CUDA platform on the execution efficient when the amount of DNA fragment has sharp increased. The improvement of DNA sequence reconstruction is more than 20 times than the previous experiment and it proves the idea is practical and efficient method.

\section{conclusion}
  This paper show a method of accelerating of reconstruction of DNA using suffix array by CUDA. The implementation was given as execute on  Intel(R) Core(TM)
i3-3110K quad-core and NVIDIA GeForce 610M GPU(kepler
architecture) which are applicable for accelerating
DNA sequence reconstruction.\\\\

   The result reveals the DNA sequence reconstruction is high time-consuming project with the growing number of DNA sequence, and construction of suffix array has been parallelized could dramatically increase the efficiency which has been speed up more than 20 times compared with the CPU series execution.

\bibliography{DNA}

\begin{thebibliography}{10}
\providecommand{\url}[1]{#1}
\csname url@samestyle\endcsname
\providecommand{\newblock}{\relax}
\providecommand{\bibinfo}[2]{#2}
\providecommand{\BIBentrySTDinterwordspacing}{\spaceskip=0pt\relax}
\providecommand{\BIBentryALTinterwordstretchfactor}{4}
\providecommand{\BIBentryALTinterwordspacing}{\spaceskip=\fontdimen2\font plus
\BIBentryALTinterwordstretchfactor\fontdimen3\font minus
  \fontdimen4\font\relax}
\providecommand{\BIBforeignlanguage}[2]{{%
\expandafter\ifx\csname l@#1\endcsname\relax
\typeout{** WARNING: IEEEtran.bst: No hyphenation pattern has been}%
\typeout{** loaded for the language `#1'. Using the pattern for}%
\typeout{** the default language instead.}%
\else
\language=\csname l@#1\endcsname
\fi
#2}}
\providecommand{\BIBdecl}{\relax}
\BIBdecl

\bibitem{Lu:Thesis:2004}
J.~P. Lu, ``{DNA Sequence Assembly},'' Master's thesis, National Chi Nan
  University, 2004.

\bibitem{kirk2010programming}
D.~B. Kirk and W.~H. Wen-mei, \emph{Programming massively parallel processors:
  a hands-on approach}.\hskip 1em plus 0.5em minus 0.4em\relax Morgan Kaufmann,
  2010.

\bibitem{anderson2008general}
J.~A. Anderson, C.~D. Lorenz, and A.~Travesset, ``General purpose molecular
  dynamics simulations fully implemented on graphics processing units,''
  \emph{Journal of Computational Physics}, vol. 227, no.~10, pp. 5342--5359,
  2008.

\bibitem{ryoo2008optimization}
S.~Ryoo, C.~I. Rodrigues, S.~S. Baghsorkhi, S.~S. Stone, D.~B. Kirk, and
  W.-m.~W. Hwu, ``Optimization principles and application performance
  evaluation of a multithreaded gpu using cuda,'' in \emph{Proceedings of the
  13th ACM SIGPLAN Symposium on Principles and practice of parallel
  programming}.\hskip 1em plus 0.5em minus 0.4em\relax ACM, 2008, pp. 73--82.

\bibitem{nvidia2007compute}
C.~Nvidia, ``Compute unified device architecture programming guide,'' 2007.

\bibitem{wen2011gpu}
W.~H. Wen-mei, \emph{GPU Computing Gems Emerald Edition}.\hskip 1em plus 0.5em
  minus 0.4em\relax Access Online via Elsevier, 2011.

\bibitem{huang1999cap3}
X.~Huang and A.~Madan, ``Cap3: A dna sequence assembly program,'' \emph{Genome
  research}, vol.~9, no.~9, pp. 868--877, 1999.

\bibitem{burks1994dna}
C.~Burks, ``Dna sequence assembly,'' \emph{Engineering in Medicine and Biology
  Magazine, IEEE}, vol.~13, no.~5, pp. 771--773, 1994.

\bibitem{parsons1993genetic}
R.~J. Parsons, S.~Forrest, and C.~Burks, ``Genetic algorithms for dna sequence
  assembly.'' in \emph{ISMB}, 1993, pp. 310--318.

\bibitem{kececioglu1995combinatorial}
J.~D. Kececioglu and E.~W. Myers, ``Combinatorial algorithms for dna sequence
  assembly,'' \emph{Algorithmica}, vol.~13, no. 1-2, pp. 7--51, 1995.

\bibitem{satish2009designing}
N.~Satish, M.~Harris, and M.~Garland, ``Designing efficient sorting algorithms
  for manycore gpus,'' in \emph{Parallel \& Distributed Processing, 2009. IPDPS
  2009. IEEE International Symposium on}.\hskip 1em plus 0.5em minus
  0.4em\relax IEEE, 2009, pp. 1--10.

\bibitem{sintorn2008fast}
E.~Sintorn and U.~Assarsson, ``Fast parallel gpu-sorting using a hybrid
  algorithm,'' \emph{Journal of Parallel and Distributed Computing}, vol.~68,
  no.~10, pp. 1381--1388, 2008.

\bibitem{boyer2008automated}
M.~Boyer, K.~Skadron, and W.~Weimer, ``Automated dynamic analysis of cuda
  programs,'' in \emph{Third Workshop on Software Tools for MultiCore Systems},
  2008.

\bibitem{karkkainen2006linear}
J.~K{\"a}rkk{\"a}inen, P.~Sanders, and S.~Burkhardt, ``Linear work suffix array
  construction,'' \emph{Journal of the ACM (JACM)}, vol.~53, no.~6, pp.
  918--936, 2006.

\end{thebibliography}
\bibliographystyle{IEEEtran}

\end{document}